%% file: main.tex
\documentclass[a4paper, conference]{IEEEtran}
\IEEEoverridecommandlockouts
\usepackage[utf8]{inputenc}
\usepackage{amsmath,amsfonts}
\usepackage[ruled,vlined]{algorithm2e}
\usepackage{algpseudocode}
\usepackage{booktabs} 
\usepackage{array}    
\usepackage{tabularx}
\usepackage[caption=false,font=normalsize,labelfont=sf,textfont=sf]{subfig}
\usepackage{textcomp}
\usepackage{stfloats}
\usepackage{url}
\usepackage{verbatim}
\usepackage{graphicx}
\usepackage{cite}
\usepackage{graphicx}
\usepackage{tikz}
\usepackage{pgfplots}
\usepackage{psfrag}
\usepackage[justification=centering]{caption}
\usetikzlibrary{arrows,snakes,shapes,shapes.geometric,fit,positioning,arrows,decorations.markings,calc}
\usepackage{mathtools}
\usepackage{amsbsy}
\usepackage{fixmath}
\usepackage{amssymb}
\usepackage{mathrsfs}
\usepackage{fancyhdr}
\usepackage{xcolor}
\usepackage{colortbl}
\usepackage{hyperref}
\newcommand{\tc}{\text{c}}
\newcommand{\tp}{\text{p}}

\newcommand{\HH}{\mathrm{H}}
\newcommand{\TT}{\mathrm{T}}
\newcommand{\tdl}{\mathrm{dl}}
\newcommand{\bs}[1]{\boldsymbol{#1}}

\DeclareMathOperator*{\argmin}{arg\,min}

\newcommand{\copyrighttext}{%
\footnotesize\textcopyright This work has been submitted to the IEEE for possible publication. Copyright may be transferred without notice, after which this version may no longer be accessible.}

\newcommand{\copyrightnotice}{%
\begin{tikzpicture}[remember picture,overlay]
\node[anchor=south,yshift=10pt] at (current page.south) {\fbox{\parbox[]{\dimexpr\textwidth-\fboxsep-\fboxrule\relax}{\copyrighttext}}};
\end{tikzpicture}%
}

\makeatletter
\def\endthebibliography{%
	\def\@noitemerr{\@latex@warning{Empty `thebibliography' environment}}%
	\endlist
}
\makeatother

\begin{document}
	
	\title{An Efficient Rate Splitting Precoding Approach
 \\in Multi-User MISO FDD Systems
	}
	
	\author{\IEEEauthorblockN{Donia Ben Amor, Michael Joham, Wolfgang Utschick }
		\IEEEauthorblockA{ 	\textit{Chair of Methods of Signal Processing} \\
			\textit{School of Computation, Information and Technology, Technical University of Munich} \\
			Munich, Germany \\
			Email: \{donia.ben-amor, joham, utschick\}@tum.de}
	}
	\maketitle
\copyrightnotice
	\begin{abstract}
In this work, we develop an efficient precoding strategy for a multi-user multiple-input-single output (MU MISO) system operating in frequency-division-duplex (FDD) mode, where rate splitting multiple access (RSMA) is implemented. To this end, we consider one-layer RS and show its significant impact on the system performance, specifically in the case where the channel state information (CSI) is incomplete at the transmitter. Based on a lower bound on the achievable rate that takes into account the CSI errors, we establish an augmented weighted average mean squared error (AWAMSE) algorithm for the RS setup denoted by AWAMSE-RS, where even the updates for the common and the private precoders are computed via analytical expressions, hence circumventing the need for interior-point methods. Simulation results validate the efficiency of our approach in terms of computational time and its competitiveness in terms of the achievable system throughput compared to state-of-the-art methods and non-RS setups.
	\end{abstract}
	
	
	\section{Introduction}
 Rate splitting multiple access (RSMA) has gained a lot of interest in the past few years due to its ability to enhance the system throughput by combining the advantages of conventional broadcasting and multicasting \cite{RSMA-6G}, \cite{RSMA-Fundamentals}, \cite{RS-PHY}, \cite{RS-PHY2}. Its benefits were shown for scenarios with perfect channel state information (CSI) at the transmitter, e.g., in \cite{RS-perfectCSI1}, \cite{RS-perfectCSI2}, as well as for more challenging scenarios with imperfect CSI \cite{RS-imperfectCSI1}, \cite{RS-imperfectCSI2}. Specifically, downlink (DL) RS is based on splitting the user's message into a private part and a common part that is decoded by all users, and therefore, it allows the users to partially decode the interference and partially treat it as noise. The potential of RS to enhance the degrees of freedom can be exploited by optimizing the resource allocation and transmission strategy \cite{RSMA-Fundamentals}.
 
 In this work, we focus on the challenging case of a frequency-division-duplex (FDD) multi-user (MU) multiple-input-single-output (MISO) system, where the transmitter (base station) has only imperfect CSI. In FDD systems, there is no channel reciprocity between the uplink (UL) and DL (as it is usually assumed in time-division-duplex (TDD) systems). Hence, the transmitter has to explicitly estimate the DL channels by sending pilots to the users and using their feedback to gain channel knowledge. Because of the dimensionality bottleneck, especially when it comes to a massive number of antennas at the transmitter, we assume that the number of DL pilots is limited and only incomplete CSI is thus available. In our previous work \cite{ICC24}, we proposed a low-complexity algorithm for designing a precoding strategy for such systems. The approach was based on first deriving a lower bound on the training-based signal-to-interference-noise-ratio (SINR) that takes into account the channel estimation errors and then establishing a connection to an average mean squared error (MSE). This allowed us to reformulate the sum rate maximization problem as an augmented weighted average MSE (AWAMSE) minimization problem that can be solved in an alternating fashion, where all the updates are given in analytical form. While this approach was shown to be efficient and competitive with state-of-the-art methods, the gap to the perfect CSI case was significant, especially when the number of users exceeded the number of pilots used for channel training. The previously mentioned advantages of RS motivated us to investigate its potential for such a challenging scenario with incomplete CSI. In order to reduce the user's hardware complexity related to successive interference cancellation (SIC), we consider the simple one-layer RS, where only one common message has to be decoded by all users, and thus, only one layer of SIC is needed at each receiver.

 Our contributions can be summarized as follows: We start by formulating the lower bounds on the training-based common and private SINRs, taking into account the channel estimation errors. Then, by relating the derived SINR expressions to the common and private average MSEs, we formulate the AWAMSE minimization problem for the RS-based system. To circumvent the need for the usually complex interior-point methods, we propose a heuristic approach to solve the underlying optimization problem. The most challenging parts when optimizing the precoders are the multicasting problem related to finding the common precoder and the minimum/maximum common rate/AWAMSE constraint. To this end, we propose to perform a search among the users to find the user index leading to the best objective. Note that all the precoder updates that need to be performed are given in closed form. Hence, the algorithm is efficient, as it will be presented later. Simulation results regarding the run time and the convergence behavior show the robustness of our method. Furthermore, the achievable system throughput is comparable to the state-of-the-art approaches proposed in \cite{RS-imperfectCSI1} and\cite{Clerckx_RS3}. The latter is based on reformulating the achievable sum rate maximization problem as a weighted MSE minimization problem and exploiting the sample average approximation (SAA) and the proposed method is hence a stochastic iterative weighted minimum MSE (SIWMMSE) approach adapted for the RS setup. \cite{RS-imperfectCSI1} uses a lower bound similar to the one we derive in the sequel but proposes an alternating procedure to solve the underlying maximization problem based on semi-definite relaxation (SDR) and concave-convex-procedure (CCCP).
	
\par This paper is organized as follows: In Section~\ref{sec:SysModel}, we present the system model for the considered setup and state the lower bounds on the training-based common and private SINRs. In Section~\ref{sec:AWAMSE}, we formulate the AWAMSE minimization problem for the RS setup and later in Section \ref{sec:PrecDesign}, we propose our solution approach. The performance of our algorithm compared to state-of-the-art methods is then evaluated through numerical simulations in Section~\ref{sec:results}.

\section{System Model}\label{sec:SysModel}
This work considers the DL of a MU MISO system, where the base station (BS) is equipped with $M$ antennas and serves $K$ single-antenna users. The channel between the BS and some generic user $k$ is denoted by $\bs{h}_k\sim \mathcal{N}_\mathbb{C}(\bs{0},\bs{C}_k)$, with $\bs{C}_k$ being the channel covariance matrix. 

To acquire the CSI, the BS sends $T_\tdl$ known pilots to the users during the channel probing phase. The collection of the channel observations at the user $k$ is assumed to be fed back from to the BS (see \cite[Section II]{ICC24}) and is given by
\begin{equation}
    \bs{y}_k=\bs{\Phi}^\HH \bs{h}_k + \bs{z}_k.
\end{equation}
where $\bs{\Phi}\in \mathbb{C}^{M\times T_\tdl}$ denotes the pilot matrix with unit-norm columns and $\bs{z}_k \sim \mathcal{N}_\mathbb{C}(\bs{0},\sigma_k^2\bs{I})$ is the downlink training noise. Here, we assume that $\sigma_k^2=1/P_\tdl$, where $P_\tdl$ is the downlink transmit power.\\
In order to reduce the training overhead, especially when it comes to a massive number of BS antennas, we assume that $T_\tdl \ll M$, such that only incomplete channel knowledge is available at the BS. To this end, we model the $k$th user's channel as \cite{Hasssibi_Hochwald}
\begin{equation}
    \bs{h}_k= \hat{\bs{h}}_k + \Tilde{\bs{h}}_k \label{eq:hk}
\end{equation}
where $\Tilde{\bs{h}}_k$ is the zero-mean estimation error whose covariance matrix is denoted by $\bs{C}_{\text{err},k}=\mathbb{E}\left[\Tilde{\bs{h}}_k\Tilde{\bs{h}}_k^\HH\right]$. $\hat{\bs{h}}_k=\mathbb{E}[ \bs{h}_k| \bs{y}_k]$ denotes the LMMSE channel estimate computed based on the channel observation $\bs{y}_k$ and is given by 
\begin{equation}
\hat{\bs{h}}_k=\bs{C}_{\bs{h}_k\bs{y}_k}\bs{C}_{\bs{y}_k}^{-1}\bs{y}_k= \bs{C}_k\bs{\Phi} (\bs{\Phi}^\HH \bs{C}_k \bs{\Phi} + \sigma_k^2 \bs{I})^{-1} \bs{y}_k.
\end{equation}

In the data transmission phase, we assume that one-layer RS is applied at the BS such that the message $W_k$ intended for user $k$ is split into one common message $W_{\tc,k}$ and one private message $W_{\tp,k}$. All common messages are combined and encoded into one super-common message $s_\tc$, whereas the private messages $W_{\tp,1}, \dots, W_{\tp,K}$ are independently encoded into $s_{\tp,1},\dots, s_{\tp,K}$. Denoting by $\bs{p}_{\tp,k}$ and $\bs{p}_\tc$ the private precoder of user $k$ and the common precoder, respectively, the received signal at user $k$ is given by
\begin{equation}
    r_k=\bs{h}_k^\HH \bs{p}_\tc s_\tc+\sum_{j=1}^K \bs{h}_k^\HH \bs{p}_{\tp,j} s_{\tp,j} + n_k
\end{equation}
where $n_k \sim \mathcal{N}_\mathbb{C}(0,\sigma_k^2)$ is the additive Gaussian noise at receiver $k$, with the normalized variance $\sigma_k^2=1/P_\tdl$.

Furthermore, we assume that one layer of SIC is implemented at the receiver side. This allows each user to first decode the common message, subtract it from the received signal $r_k$, and lastly decode its own private message by treating other users' private messages as noise. The message split gives rise to two rate expressions: the common rate $R_{\tc,k}$ and the private rate $R_{\tp,k}$ for each user $k$. Due to imperfect CSI, the BS has no knowledge about the actual achievable user rates. Thus, relying on the perfect CSI rate expressions when designing the transmit strategy can lead to an overestimation of the achievable rates and, consequently, poor system performance. \\
Therefore, we reside to a lower bound on the $k$th user's rate as derived in \cite{ICC24}, which considers the imperfect channel knowledge. To this end, we rewrite the received signal at user $k$ using the model in \eqref{eq:hk} as follows
\begin{equation}
    r_k=\underbrace{\hat{\bs{h}}_k^\HH \bs{p}_\tc s_\tc}_\text{desired part} + \underbrace{\Tilde{\bs{h}}_k^\HH \bs{p}_\tc s_\tc + \sum_{j=1}^K (\hat{\bs{h}}_k^\HH\bs{p}_{\tp,j} +\Tilde{\bs{h}}_k^\HH\bs{p}_{\tp,j}) s_{\tp,j} + n_k}_\text{interference + noise}.
\end{equation}
Assuming that the data signals are independent among each other and from the noise  signal $n_k$ and that $s_\tc, s_{\tp,j}\sim \mathcal{N}_\mathbb{C}(0,1)$, the common SINR of user $k$ can be lower bounded as follows \cite{ICC24}
\begin{equation}
    \overline{\gamma}_{\tc,k} = \frac{|\hat{\bs{h}}_k^\HH \bs{p}_\tc|^2}{\bs{p}_{\tc}^\HH  \bs{C}_{\text{err},k} \bs{p}_{\tc} +\sum_{j=1}^K  \bs{p}_{\tp,j}^\HH  \bs{C}_{\text{err},k} \bs{p}_{\tp,j} + |\hat{\bs{h}}_k^\HH \bs{p}_{\tp,j}|^2+\sigma_k^2}.\label{eq:SINRc} 
\end{equation}

We assume that perfect SIC is performed, i.e., the common message is retrieved error-free. Note that this assumption is not too idealistic since each user needs to acquire only the scalar effective channel $\bs{h}_k^\HH\bs{p}_\tc$, which can be ensured by sending one precoded dedicated pilot in the downlink. We also want to emphasize that this assumption should not be confused with perfect CSI at the receivers. Even in the latter case, estimating the scalar effective channel is necessary for performing SIC since the users have no information about the used common precoder $\bs{p}_\tc$.\\
By subtracting the already decoded common part from the received signal $r_k$ and using the channel error model in \eqref{eq:hk}, the lower bound on the private SINR of user $k$ is given by
\begin{equation}
     \overline{\gamma}_{\tp,k} = \frac{|\hat{\bs{h}}_k^\HH \bs{p}_{\tp,k}|^2}{\sum_{j=1}^K  \bs{p}_{\tp,j}^\HH  \bs{C}_{\text{err},k} \bs{p}_{\tp,j} +\sum_{j\neq k}|\hat{\bs{h}}_k^\HH \bs{p}_{\tp,j}|^2+\sigma_k^2}.\label{eq:SINRp} 
\end{equation}
Hence, the common and private rate lower bounds for user $k$ are respectively given by $\overline{R}_{\tc,k}=\log_2(1+\overline{\gamma}_{\tc,k})$
and $\overline{R}_{\tp,k}=\log_2(1+\overline{\gamma}_{\tp,k})$.

Our goal is to design the common precoder $\bs{p}_\tc$ and the private precoding matrix $\bs{P}_\tp=[\bs{p}_{\tp,1},\dots, \bs{p}_{\tp,K}]$ such that the lower bound on the training-based sum rate is maximized, i.e.,
\begin{equation}\label{eq:maxSR}
    \underset{\bs{p}_\tc,\bs{P}_\tp}{\max} \sum_{j=1}^K \overline{R}_{\tp,j} + \underset{k}{\min}\: \overline{R}_{\tc,k} \quad \text{s.t.} \: \|\bs{p}_\tc\|^2 +  \|\bs{P}_\tp\|_\mathrm{F}^2  \leq 1.
\end{equation}
The inner minimization over the users in \eqref{eq:maxSR} ensures that the common message is decodable at all receivers, and the resulting common rate is thus given by the minimum common rate among all users.

The throughput maximization problem for the non-RS setup is known to be NP-hard \cite{SpecMang_Luo}, and the inner minimization over the users for the common rate in the RS setup makes it more involved. Therefore, we will formulate an augmented weighted average MSE (AWAMSE) minimization problem in the next section, which helps to reveal hidden convexity properties of \eqref{eq:maxSR} and is thus easier to tackle.\\
In a later section, we will present an efficient heuristic solution approach to solve the underlying AWAMSE minimization problem that is shown to be competitive with more computationally demanding state-of-the-art approaches.

\section{AWAMSE Minimization for RS}\label{sec:AWAMSE}
Similar to \cite{ICC24}, we want to establish a relationship between the SINR lower bounds given in \eqref{eq:SINRc} and \eqref{eq:SINRp} to average MSEs, where the averaging takes place w.r.t. the channel estimates. To this end, we introduce for each user $k$ two receive filters $g_{\tc,k}$ and $g_{\tp,k}$ for the common and the private data signals, respectively. The average common and private MSEs for user $k$ are then respectively given by
\begin{align}  \overline{\varepsilon}_{\tc,k}&=1-2\Re\{g_{\tc,k}\hat{\bs{h}}_k^\HH \bs{p}_\tc\} +|g_{\tc,k}|^2 (T_{\tc,k}+\sigma_k^2) \label{eq:AMSEc}\\
 \overline{\varepsilon}_{\tp,k}&= 1 -2\Re\{g_{\tp,k}\hat{\bs{h}}_k^\HH \bs{p}_{\tp,k}\}+ |g_{\tp,k}|^2(T_{\tp,k}+\sigma_k^2)
 \label{eq:AMSEp}
\end{align}
where $T_{\tc,k}=|\hat{\bs{h}}_k^\HH \bs{p}_\tc|^2+\bs{p}_\tc^\HH \bs{C}_{\text{err},k}\bs{p}_\tc+ T_{\tp,k}$ and $T_{\tp,k}~=~\sum_j |\hat{\bs{h}}_k^\HH \bs{p}_{\tp,j}|^2 +\bs{p}_{\tp,j}^\HH \bs{C}_{\text{err},k}\bs{p}_{\tp,j}$.\\
Using the MSE minimizing filters $g_{\tc,k}^\text{MMSE}$ and $g_{\tp,k}^\text{MMSE}$ given by
\begin{align}
    g_{\tc,k}^\text{MMSE}&=\bs{p}_\tc^\HH \hat{\bs{h}}_k (T_{\tc,k}+\sigma_k^2)^{-1},\\
     g_{\tp,k}^\text{MMSE}&=\bs{p}_{\tp,k}^\HH \hat{\bs{h}}_k (T_{\tp,k}+\sigma_k^2)^{-1}
\end{align}
we have the following relationship between the rate lower bounds and the AMSEs
\begin{equation}
    \overline{R}_{\tc,k}=-\log_2(\overline{\varepsilon}_{\tc,k}^\text{MMSE}), \quad \text{and}\quad \overline{R}_{\tp,k}=-\log_2(\overline{\varepsilon}_{\tp,k}^\text{MMSE}).
\end{equation}
Next, we additionally define an augmented weighted average MSE (AWAMSE) for the common and private signals for each user $k$ 
\begin{equation}   \overline{\xi}_{\tc,k}=u_{\tc,k}\overline{\varepsilon}_{\tc,k}-\log_2(u_{\tc,k}), \quad \overline{\xi}_{\tp,k}=u_{\tc,k}\overline{\varepsilon}_{\tp,k}-\log_2(u_{\tp,k})
\end{equation}
where we introduced the weighting factors $u_{\tc,k}$ and $u_{\tp,k}$. For fixed precoders and receive filters, the optimal common and private weights are obtained by setting the first derivative of $\overline{\xi}_{\tc,k}$ and $\overline{\xi}_{\tp,k}$ to zero, which yields $u_{\tc,k}^\text{MMSE}=1/\overline{\varepsilon}_{\tc,k}^\text{MMSE}$ and $u_{\tp,k}^\text{MMSE}=1/\overline{\varepsilon}_{\tp,k}^\text{MMSE}$. This leads to the AWAMSE-rate relationships $\overline{\xi}_{\tc,k}^\text{MMSE}=1-\overline{R}_{\tc,k}$ and $\overline{\xi}_{\tp,k}^\text{MMSE}=1-\overline{R}_{\tp,k}$.

Using the established connections between the rate lower bounds and the AWAMSEs, we can formulate the sum rate maximization problem in \eqref{eq:maxSR} as the following minimization problem
\begin{equation}\label{eq:minAWAMSE}
    \underset{\substack{\bs{p}_\tc,\bs{P}_\tp,\\ \{g_{\tc,i}\},\{g_{\tp,i}\}, \\ \{u_{\tc,i}\},\{u_{\tp,i}\}}}{\min} \sum_{j=1}^K \overline{\xi}_{\tp,j} + \underset{k}{\max}\,\overline{\xi}_{\tc,k} \quad \text{s.t.} \quad 
   \|\bs{p}_\tc\|^2 + \|\bs{P}_\tp\|_\mathrm{F}^2 \leq 1.
\end{equation}
Note that for fixed receive filters $\{g_{\tc,i}\},\{g_{\tp,i}\}$ and weights $\{u_{\tc,i}\},\{u_{\tp,i}\}$, the optimization problem in \eqref{eq:minAWAMSE} is a quadratically constrained quadratic program w.r.t. the precoders and can be solved using interior-point solvers. However, this can be very costly, especially since the solver has to be re-run for each new update of the receive filters and the weights. Next, we will present an efficient heuristic approach to solve \eqref{eq:minAWAMSE}.

\section{Precoder Design for AWAMSE RS}\label{sec:PrecDesign}
In \cite{ICC24}, a closed-form solution for the precoder update was presented when the weights and receive filters were fixed, whereby no RS was implemented. A similar derivation will be conducted in the following to adapt the proposed approach for the RS setup.

Before we proceed to solve \eqref{eq:minAWAMSE}, let us introduce a common scaling of the receive filters denoted by $\beta^{-1}$ (see \cite{WienerFilter}), which yields $\Tilde{g}_{\tc,k}=\beta^{-1} g_{\tc,k}$ and $\Tilde{g}_{\tp,k}=\beta^{-1}{g}_{\tp,k}$. Accordingly, the precoding vectors have to be scaled with $\beta$.\\ By defining the overall precoding matrix $\bs{P}=[\bs{p}_\tc, \bs{P}_\tp]$, the transmit power constraint reads as $\|\bs{P}\|_\mathrm{F}^2\leq 1$. \\
Assuming that the scaled precoding matrix $\Tilde{\bs{P}}=\beta \bs{P}$ is optimal \cite{WienerFilter}, and from the equality of the power constraint in the optimum, we can infer an expression for $\beta$, i.e.,
\begin{equation}
\|\Tilde{\bs{P}}\|_\mathrm{F}^2=\beta^2\|\bs{P}\|_\mathrm{F}^2=1 \Rightarrow   \beta^{-1}=\|\bs{P}\|_\mathrm{F}.
\label{eq:ScaleBeta}
\end{equation}
With this choice of $\beta$, the transmit power constraint is satisfied with equality, \eqref{eq:minAWAMSE} can be reformulated as the following unconstrained optimization problem (\cite{RethinkWMMSE}) 
\begin{equation}\label{eq:AWAMSE_RS2}
    \begin{aligned}
    \underset{\substack{\bs{P},\\ \bs{G},
    \\ \bs{U}}}{\min}
    \sum_{j=1}^K &u_{\tp,j} \Big[ 1-2\Re\{g_{\tp,j}\hat{\bs{h}}_j^\HH \bs{p}_{\tp,j}\}  +  |g_{\tp,j}|^2 
   \Big. \big(\sigma_j^2\|\bs{P}\|_\mathrm{F}^2 +  T_{\tp,j}\big) \Big] \\- & \log_2(u_{\tp,j}) \\
+ \underset{k}{\max}\, & u_{\tc,k}  \Big[1-2\Re\{g_{\tc,k}\hat{\bs{h}}_k^\HH \bs{p}_\tc\}+   |g_{\tc,k}|^2 \big(\sigma_k^2\|\bs{P}\|_\mathrm{F}^2 + T_{\tc,k}  \big) \Big] \\ - &\log_2(u_{\tc,k})
    \end{aligned}
\end{equation}
where we collected the receive filters in $\bs{G}=[[g_{\tc,1},\dots,g_{\tc,K}]^\TT,[g_{\tp,1},\dots,g_{\tp,K}]^\TT]$ and the weights in $\bs{U}=[[u_{\tc,1},\dots,u_{\tc,K}]^\TT,[u_{\tp,1},\dots,u_{\tp,K}]^\TT]$.
Due to the inner maximization problem over the users and the multicast problem related to the common precoder design, \eqref{eq:AWAMSE_RS2} is still involved, even for fixed weights and receive filters.

To tackle this challenging problem, we propose the following heuristic approach. For fixed receive filters and weights, we first fix the user index for the common AWAMSE to some $k_\tc$ and drop the inner maximization over the users. Therefore, the corresponding optimization problem reads as
\begin{equation}\label{eq:AWAMSE_RS3}
    \begin{aligned}
    \underset{\bs{P}}{\min}\,
    &\sum_{j=1}^K u_{\tp,j} \Big[ 1-2\Re\{g_{\tp,j}\hat{\bs{h}}_j^\HH \bs{p}_{\tp,j}\}  +  |g_{\tp,j}|^2 
   \Big. \big(\sigma_j^2\|\bs{P}\|_\mathrm{F}^2 +  T_{\tp,j}\big) \Big] \\- & \log_2(u_{\tp,j}) \\
+ & u_{\tc,k_{\tc}}  \Big[1-2\Re\{g_{\tc,k_{\tc}}\hat{\bs{h}}_{k_{\tc}}^\HH \bs{p}_\tc\}+   |g_{\tc,k_{\tc}}|^2 \big(\sigma_{k_{\tc}}^2 \|\bs{P}\|_\mathrm{F}^2 + T_{\tc,k_{\tc}}  \big) \Big] \\ - &\log_2(u_{\tc,k_{\tc}}).
    \end{aligned}
\end{equation}
Consequently, each choice of $k_\tc \in \{1,\dots, K\}$ leads to closed-form expressions for the common and private precoders. These are obtained by simply setting the first derivative of the objective of \eqref{eq:AWAMSE_RS3} w.r.t. $\bs{p}_\tc$ and $\bs{p}_{\tp,j}$ to zero. The corresponding precoder expressions are given as a function of the chosen user index $k_\tc$
\begin{align}
    \bs{p}_\tc(k_\tc)&=\bigg(\bs{A}+ \bs{B} \bigg)^{-1}u_{\tc,k_\text{c}}g_{\tc,k_\tc}^* \hat{\bs{h}}_{k_\tc} \label{eq:pckmax}\\
     \bs{p}_{\tp,j}(k_\tc)&=\bigg(\bs{A}+ \bs{B} + \bs{C} \bigg)^{-1}u_{\tp,j}g_{\tp,j}^* \hat{\bs{h}}_{j} \label{eq:pj}
\end{align}
where we introduced the matrices 
\begin{align}
\bs{A}&=u_{\tc,k_\tc}|g_{\tc,k_\tc}|^2\big(\hat{\bs{h}}_{k_\tc}\hat{\bs{h}}_{k_\tc}^\HH + \bs{C}_{\text{err},k_\tc} +\sigma_{k_\tc}^2 \bs{I}\big)\\
    \bs{B}&=\sum_{i=1}^K u_{\tp,i}|g_{p,i}|^2 \sigma_i^2 \bs{I} \\
    \bs{C}&=\sum_{i=1}^K u_{\tp,i}|g_{p,i}|^2 \big( \hat{\bs{h}}_i \hat{\bs{h}}_i^\HH + \bs{C}_{\text{err},i} \big).
\end{align}
For every possible $k_\tc\in \{1,\dots,K\}$, we obtain a precoding matrix denoted by $\bs{P}(k_\tc)$ and the overall objective can be evaluated according to
\begin{equation} \label{eq:xiTot}
    \xi(k_\tc)=\sum_j \xi_{\tp,j}(\bs{P}(k_\tc))+ \underset{k}{\max}\, \xi_{\tc,k}(\bs{P}(k_\tc)).
\end{equation}
If the update of the precoders for every possible user index does not improve the objective compared to the previous iteration, the algorithm stops, and the precoder from the previous iteration is the final solution. Otherwise, the user index $k_{\min}$ leading to the smallest objective among all possible user indices is selected, i.e., $k_{\min}=\underset{k_\tc}{\argmin}\,\xi(k_\tc)$,
and the precoder is updated using the solution obtained for user index $k_{\min}$. Note that the decodability of the common message is ensured in each update since we consider the overall objective based on the maximum common MSE across all users for the selected precoders. \\
Once the precoder is fixed, the receive filters $g_{\tc,k}$ and $g_{\tp,k}$ and the weights $u_{\tc,k}$ and $u_{\tp,k}$ can be computed according to the following equations
\begin{align}
    g_{\tc,k}&= \bs{p}_\tc^\HH \hat{\bs{h}}_k (T_{\tc,k}+\sigma_k^2\|\bs{P}\|_\text{F}^2)^{-1} \label{eq:gckScaled} \\
    g_{\tp,k}&= \bs{p}_{\tp,k}^\HH \hat{\bs{h}}_k  (T_{\tp,k}+\sigma_k^2\|\bs{P}\|_\text{F}^2)^{-1} \label{eq:gpkScaled}
    \end{align}
    \begin{align}
    u_{\tc,k}&=\left(1 - |\hat{\bs{h}}_k^\HH \bs{p}_\tc|^2 (T_{\tc,k}+\sigma_k^2\|\bs{P}\|_\text{F}^2)^{-1}\right)^{-1} \label{eq:uckScaled}\\
    u_{\tp,k}&=\left(1 - |\hat{\bs{h}}_k^\HH \bs{p}_{\tp,k}|^2 (T_{\tp,k}+\sigma_k^2\|\bs{P}\|_\text{F}^2)^{-1}  \right)^{-1}\label{eq:upkScaled}.
\end{align}
This procedure is repeated until the absolute change of the objective drops below a certain threshold or the maximum number of iterations is reached. Finally, the precoding matrix must be re-scaled to satisfy the power constraint (cf. \eqref{eq:ScaleBeta}).\\
The corresponding AWAMSE-RS algorithm is summarized in Algorithm~\ref{alg:myalgorithm}.
\begin{algorithm}
\caption{AWAMSE-RS Algorithm}
\label{alg:myalgorithm}
    Initialize $\bs{P}\leftarrow \bs{P}^{(0)}$, $\xi_{\min}\leftarrow \infty$ and $n\leftarrow 1$ \\
    \While{No convergence is reached}{
    Update $g_{\tc,k}$ and $g_{\tp,k}$ according to \eqref{eq:gckScaled} and \eqref{eq:gpkScaled}\\
    Compute $u_{\tc,k}$ and $u_{\tp,k}$ according to \eqref{eq:uckScaled} and \eqref{eq:upkScaled}\\
    \For{$k_\tc=1$ to $K$}{
    Determine $\bs{P}(k_\tc)$ using \eqref{eq:pckmax} and \eqref{eq:pj}\\
    Compute $\xi(k_\tc)\leftarrow \sum_j \xi_{\tp,j}+\underset{k}{\max}\, \xi_{\tc,k}$}
   \If{$\underset{k_\tc}{\min}\, \xi(k_\tc)>\xi_{\min}$}{$\bs{P}\leftarrow \bs{P}^{(n-1)}$ \\ break} \SetAlgoNlRelativeSize{-1}
   \Else{$k_{\min}\leftarrow \underset{k_\tc}{\argmin}\,\xi(k_\tc)$ \\ $\xi_{\min}\leftarrow  \xi(k_{\min})$\\ $\bs{P}^{(n)}\leftarrow \bs{P}(k_{\min})$}
  $n\leftarrow n+1$} 
   Scale $\bs{P}$ with $1/\|\bs{P}\|_\mathrm{F}$ to satisfy the power constraint
\end{algorithm}

\section{Numerical Results}\label{sec:results}
We evaluate the performance of the proposed algorithm for a system with $M=16$ BS antennas and $K=5$ users and varying numbers of training pilots. The results are averaged over $100$ channel realizations, where the channel of some generic user $k$ is generated according to $\bs{h}_k \sim \mathcal{N}_\mathbb{C}(\bs{0},\bs{C}_k)$. $\bs{C}_k$ is obtained from a randomly selected component of a Gaussian mixture model (GMM) that is fitted to the DL training observations of channels generated using QUAsi Deterministic RadIo channel GenerAtor (QuaDRiGa) \cite{GMM2}. 

Since it has been shown in \cite{ZF_LS} that for $T_\tdl\geq K$, full degrees of freedom are attainable at asymptotically high transmit powers even with incomplete CSI, we focus on the more challenging case, where $T_\tdl<K$. In this case, RS has a high potential to increase the system's degrees of freedom in the high-power regime, as observed in the results later.

As a benchmark for the performance of our RS algorithm, we show the results obtained with two alternating optimization methods, namely the stochastic IWMMSE (SIWMMSE) approach proposed in \cite{Clerckx_RS3} for the RS setup (denoted here by SIWMMSE-RS) and the semi-definite relaxation (SDR) approach presented in \cite{RS-imperfectCSI1} (denoted by SDR-RS).\\
In the former, the authors suggest reformulating the ergodic sum rate maximization problem as an augmented weighted MSE minimization problem by introducing receive filters and weights for each receiver. Due to the stochastic nature of the underlying optimization problem, sample average approximation (SAA) is used to find the optimal weights and receive filters inside every iteration by averaging over several samples of the channel. The samples are generated based on the channel estimate and the known channel statistics. Once the weights and receive filters are fixed, interior-point methods are used to solve for the precoders. In our simulations, we employ CVX \cite{cvx} as an optimization toolbox, as suggested in \cite{Clerckx_RS3}. \\
\cite{RS-imperfectCSI1} proposed to solve the maximization problem of the sum rate lower bound via an alternating procedure. To this end, the authors suggest introducing a set of auxiliary variables and applying SDR and CCCP to solve for the precoders and the auxiliary variables. Due to SDR, the optimal precoders that deliver the best objective have to be found by performing a random vector search.\\
Note that we do not assume Rayleigh fading channels in contrast to both works. Instead, they are assumed to exhibit full diagonal covariance matrices and not just a scaled identity. 

Additionally, we compare our results to the non-RS approach proposed in our recent work \cite{ICC24} and to the MMSE precoder from \cite{MMSEprecoder}. The latter is given by $\bs{P}^\text{mmse}=\delta \left(\hat{\bs{H}}\hat{\bs{H}}^\HH+\bs{C}_\text{err}+\eta \bs{I}\right)^{-1}\hat{\bs{H}}$, where $\delta$ is used for normalization, $\bs{C}_\text{err}=\sum_k \bs{C}_{\text{err},k}$ and $\eta=M/P_\tdl$ is a regularization parameter. 

For our AWAMSE-RS approach and the SIWMMSE-RS method proposed in \cite{Clerckx_RS3}, we initialize the common precoder with $\alpha_\tc \bs{u}$, where $\bs{u}$ is the left singular vector of $\hat{\bs{H}}$ corresponding to the largest singular value and $\alpha_\tc \in [0,1]$ is the power fraction allocated to the common stream transmission. The private precoding matrix is initialized with the normalized MMSE precoder, i.e., $\bs{P}_\tp=(1-\alpha_\tc) \bs{P}^\text{mmse}/\| \bs{P}^\text{mmse}\|_\mathrm{F}$. As for the SDR-RS approach from \cite{RS-imperfectCSI1}, we randomly initialize the auxiliary variables as proposed by the authors and use $N_\text{r}=1000$ random vectors in the randomization step.

\begin{figure}[!ht]
	\centering
 \scalebox{0.45}{\input{SR_M=16_K=5_Tdl=3_alpha=0.5_Quadriga}}
	\caption{Achievable SR versus $P_\tdl$ for $M=16$ antennas, $K=5$ users and $T_\tdl=3$ pilots using different precoding strategies} 
	\label{fig:SR_Tdl=3}
\end{figure}
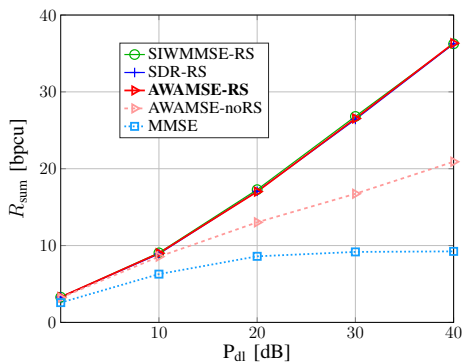
In Fig.~\ref{fig:SR_Tdl=3}, we plot the average achievable sum rate in bits per channel use (bpcu) versus the DL transmit power for the different precoding strategies. Here, we consider a setup with $M=16$ antennas, $K=5$ users, and $T_\tdl=3$ training pilots. The common power fraction $\alpha_\tc$ is set to 0.5. One can easily observe that all the RS approaches exhibit almost the same performance over the whole power range. Furthermore, they outperform the non-RS approaches, and the gap becomes significantly larger with increasing transmit power. While the two non-RS approaches, MMSE and AWAMSE-noRS, exhibit high SNR slopes of 0.005 and 0.47, respectively, the RS approaches show a high SNR slope of approximately \textbf{0.97}. This shows that RS can boost the performance of such a system where the channel knowledge is limited at the BS by increasing the degrees of freedom, especially in the high-power regime. \\
Now that we have seen that our proposed algorithm is competitive with its counterpart approaches in terms of the system throughput, we will assess its performance in terms of the average run time needed until convergence compared to these methods. The corresponding results are shown in Table~\ref{tab:runtimes} for different transmit powers. Since the SIWMMSE-RS and SDR-RS approaches rely on interior-point methods to solve the underlying optimization problem and no analytical solutions are used, the run times of these algorithms are remarkably higher (by a factor of at least 100) than our proposed approach.

\begin{table}[!ht]
\centering
    \begin{tabular}{|c|c|c|c|c|}
    
    \hline
\rowcolor{lightgray}  \hfill &0\,dB & 20\,dB &40\,dB \\
  \hline
SIWMMSE-RS & 21.9s & 111.6s  & 96.6s \\
\hline
SDR-RS & 62.9s  & 82.23s &  116.1s  \\
\hline
\textbf{AWAMSE-RS} & \textbf{0.033}s  &  \textbf{0.094}s   & \textbf{0.143}s \\
  \hline
\end{tabular}
\caption{Average run times in seconds (s) of the RS algorithms for $M=16$ antennas, $K=5$ users, $T_\tdl=3$ pilots and different transmit powers}
    \label{tab:runtimes}
\end{table}

\begin{figure}[!ht]
\begin{minipage}{0.48\columnwidth}
\centering
\scalebox{.4}{\input{CDF_Niter_M=16_Tdl=3_Pdl=20dB_alpha=0.5}}
\caption{CDF plot of the number of iterations needed until convergence \\ for $P_\tdl=20$~dB}
\label{fig:CDFniter}
\end{minipage}
\begin{minipage}{0.1\columnwidth}
\end{minipage}
\begin{minipage}{0.48\columnwidth}
 \centering
\scalebox{.4}{\input{CDF_Runtime_M=16_Tdl=3_Pdl=20dB_alpha=0.5}}
\caption{CDF plot of the run times needed until convergence \\for $P_\tdl=20$~dB}
 \label{fig:CDFruntime}
\end{minipage}
\end{figure}
Figs.~\ref{fig:CDFniter} and \ref{fig:CDFruntime} display respectively the cumulative distribution function (CDF) plots of the number of iterations and of the run times needed until convergence for the three RS approaches at $P_\tdl=20$~dB. While our proposed approach might exhibit more iterations than the SDR-RS method to reach convergence, as can be seen in Fig.~\ref{fig:CDFniter}, one can easily observe in Fig.~\ref{fig:CDFruntime} that the run time of the AWAMSE-RS algorithm is negligible compared to the other two approaches which highlights the efficiency of our approach in terms of the computational complexity.

\begin{figure}[!ht]
	\centering
 \scalebox{0.45}{\input{SR_M=16_K=5_Tdl=2_alpha=0.5_Quadriga}}
	\caption{Achievable SR versus $P_\tdl$ for $M=16$ antennas, $K=5$ users and $T_\tdl=2$ pilots using different precoding strategies} 
	\label{fig:SR_Tdl=2}
\end{figure}
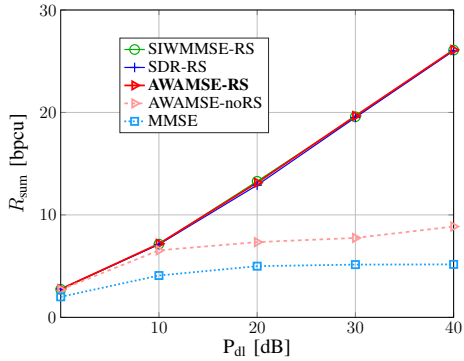

In Fig.~\ref{fig:SR_Tdl=2}, we again plot the average achievable sum rate for the same setup except for the number of pilots that is now set to $T_\tdl=2$, leading to a more challenging scenario. Here again, all RS approaches deliver approximately the same throughput. The advantage of applying RS becomes even more visible compared to the previous case. At $20$~dB, the gap of the RS methods to AWAMSE-noRS is 5.8~bpcu and reaches approximately 17~bpcu at 40~dB.

Finally, we consider the power allocation of the RS approaches at $40$~dB among the common and private streams for a system with $M=16$ antennas and $K=5$ users. In our recent work \cite{ZF_LS}, we have shown that the number of active users that can be served is upper bounded by $\min(T_\tdl, K)$ in a non-RS setup. This observation is confirmed also in this RS-setup by Figs.~\ref{fig:PowAlloc2pilots} and~\ref{fig:PowAlloc3pilots} where $T_\tdl=2$ and $T_\tdl=3$, respectively. The common power fraction is denoted by "Com" and user $k$ is denoted by UE$k$, $k=1, \dots, K$. One can see that the power allocation pattern is basically the same for all algorithms. Apart from the common stream, only $T_\tdl$ users are active, which shows that all methods converge to the same power allocation strategy. 

\begin{figure}[!ht]
\begin{minipage}{0.48\columnwidth}
\centering
\scalebox{.4}{\input{PowerAlloc_M=16_Tdl=2_Pdl=40dB_alpha=0.5}}
\caption{Power allocation for $T_\tdl=\textbf{2}$ \textbf{pilots}}
\label{fig:PowAlloc2pilots}
\end{minipage}
\begin{minipage}{0.1\columnwidth}
\end{minipage}
\begin{minipage}{0.48\columnwidth}
 \centering
\scalebox{.4}{\input{PowerAlloc_M=16_Tdl=3_Pdl=40dB_alpha=0.5}}
\caption{Power allocation for $T_\tdl=\textbf{3}$ \textbf{pilots}}
 \label{fig:PowAlloc3pilots}
\end{minipage}
\end{figure}

\section{Conclusion}
We have proposed an efficient solution approach for the throughput maximization in the downlink of an MU MISO system featuring one-layer RS. We have shown the advantage of applying RS to challenging scenarios with very limited channel knowledge at the transmitter through numerical simulations. Furthermore, our proposed approach was not only competitive with other state-of-the-art RS approaches, but it also exhibited considerably faster run time. This is due to the analytical expressions used for the precoder update and the absence of interior-point methods usage, which is usually time-consuming.

	\bibliographystyle{unsrt}


\end{document}

%% file: SR_M=16_K=5_Tdl=3_alpha=0.5_Quadriga.tex
%
%
\definecolor{mycolor1}{rgb}{0.07451,0.62353,1.00000}
\begin{tikzpicture}

\begin{axis}[%
width=4.521in,
height=3.566in,
at={(0.758in,0.481in)},
scale only axis,
xmin=0,
xmax=40,
xlabel style={font=\color{white!15!black},font=\LARGE},
xlabel={$\text{P}_{\text{dl}}\text{ [dB]}$},
ymin=0,
ymax=40,
ylabel style={font=\color{white!15!black}, font=\LARGE},
ylabel={$R_\text{sum}$ [bpcu]},
axis background/.style={fill=white},
xtick={10,20,30,40},
ytick={0,10,20,30,40},
xmajorgrids,
ymajorgrids,
yticklabel style={font=\Large},
xticklabel style={font=\Large},
legend style={at={(0.15,0.6)}, anchor=south west, legend cell align=left, align=left, draw=white!15!black, font=\Large}
]

\addplot [color=green!70!black, opacity=0.8, line width=1pt, mark=o, mark options={solid},mark size=4pt]
  table[row sep=crcr]{%
0	3.28103406199617\\
10	9.05209951060546\\
20	17.2775637450285\\
30	26.805747481579\\
40	36.2503039969568\\
};
\addlegendentry{SIWMMSE-RS}

\addplot [color=blue,opacity=0.8, line width=1pt, mark=+, mark options={solid, blue}, mark size=4pt]
  table[row sep=crcr]{%
0	3.29908423201762\\
10	8.9108049829936\\
20	17.0602346991281\\
30	26.4091575980621\\
40	36.2458987824203\\
};
\addlegendentry{SDR-RS}

\addplot  [color=red, line width=1.5pt, mark=triangle, mark options={solid, rotate=270, red},mark size=4pt]
  table[row sep=crcr]{%
0	3.29291868391095\\
10	8.99892666103075\\
20	17.0214094379893\\
30	26.5434311877461\\
40	36.3297336204969\\
};
\addlegendentry{ \textbf{AWAMSE-RS}}

\addplot  [color=red!40!white,line width=1.5pt, mark=triangle, mark options={solid ,rotate=270}, dashed, mark size=4pt]
  table[row sep=crcr]{%
0	3.20322396209486\\
10	8.55976300460853\\
20	13.0245227311881\\
30	16.7398375938505\\
40	20.9079989206846\\
};
\addlegendentry{AWAMSE-noRS}

\addplot [color=mycolor1,line width=1.5pt, mark=square, mark options={solid}, dotted, mark size=3]
  table[row sep=crcr]{%
0	2.57505800534359\\
10	6.28140647335208\\
20	8.59862579120115\\
30	9.17271893561829\\
40	9.24064227289357\\
};
\addlegendentry{MMSE}

\end{axis}

\end{tikzpicture}%

%% file: CDF_Niter_M=16_Tdl=3_Pdl=20dB_alpha=0.5.tex
%
%
\definecolor{mycolor1}{rgb}{0.00000,0.44700,0.74100}%
\definecolor{mycolor2}{rgb}{0.85000,0.32500,0.09800}%
\definecolor{mycolor3}{rgb}{0.92900,0.69400,0.12500}%
\begin{tikzpicture}
\begin{axis}[%
width=3.721in,
height=3.566in,
at={(0.758in,0.481in)},
scale only axis,
xmin=0,
xmax=350,
xlabel style={font=\color{white!15!black},font=\LARGE},
xlabel={Number of iterations},
ymin=0,
ymax=1,
yticklabel style={font=\LARGE},
xticklabel style={font=\LARGE},
ylabel style={font=\color{white!15!black}, font=\huge},
ylabel={CDF},
axis background/.style={fill=white},
xtick={0,100,200,300},
ytick={0,0.5,1},
xmajorgrids,
ymajorgrids,
legend style={at={(0.48,0.5)}, anchor=south west, legend cell align=left, align=left, draw=white!15!black, font=\LARGE}
]

\addplot  [color=red, line width=2.5pt]
  table[row sep=crcr]{%
4	0.01\\
4	0.02\\
4	0.03\\
5	0.04\\
5	0.05\\
5	0.06\\
5	0.07\\
6	0.08\\
6	0.09\\
6	0.1\\
6	0.11\\
6	0.12\\
6	0.13\\
7	0.14\\
7	0.15\\
7	0.16\\
7	0.17\\
7	0.18\\
7	0.19\\
7	0.2\\
7	0.21\\
8	0.22\\
8	0.23\\
9	0.24\\
9	0.25\\
9	0.26\\
9	0.27\\
9	0.28\\
9	0.29\\
9	0.3\\
9	0.31\\
9	0.32\\
9	0.33\\
10	0.34\\
10	0.35\\
10	0.36\\
10	0.37\\
11	0.38\\
11	0.39\\
11	0.4\\
11	0.41\\
11	0.42\\
12	0.43\\
13	0.44\\
13	0.45\\
13	0.46\\
14	0.47\\
15	0.48\\
15	0.49\\
15	0.5\\
15	0.51\\
15	0.52\\
16	0.53\\
16	0.54\\
16	0.55\\
17	0.56\\
17	0.57\\
17	0.58\\
18	0.59\\
19	0.6\\
20	0.61\\
20	0.62\\
22	0.63\\
22	0.64\\
22	0.65\\
22	0.66\\
23	0.67\\
23	0.68\\
24	0.69\\
24	0.7\\
25	0.71\\
27	0.72\\
28	0.73\\
29	0.74\\
33	0.75\\
33	0.76\\
34	0.77\\
35	0.78\\
36	0.79\\
41	0.8\\
42	0.81\\
42	0.82\\
46	0.83\\
47	0.84\\
48	0.85\\
48	0.86\\
49	0.87\\
51	0.88\\
52	0.89\\
53	0.9\\
61	0.91\\
63	0.92\\
65	0.93\\
69	0.94\\
70	0.95\\
76	0.96\\
79	0.97\\
82	0.98\\
95	0.99\\
103	1\\
};
\addlegendentry{\textbf{AWAMSE-RS}}

\addplot [color=blue,opacity=0.8, line width=1.5pt]table[row sep=crcr]{%
5	0.01\\
5	0.02\\
5	0.03\\
5	0.04\\
6	0.05\\
6	0.06\\
6	0.07\\
6	0.08\\
6	0.09\\
6	0.1\\
6	0.11\\
6	0.12\\
6	0.13\\
6	0.14\\
6	0.15\\
6	0.16\\
6	0.17\\
6	0.18\\
6	0.19\\
6	0.2\\
6	0.21\\
6	0.22\\
6	0.23\\
6	0.24\\
6	0.25\\
6	0.26\\
6	0.27\\
6	0.28\\
6	0.29\\
6	0.3\\
6	0.31\\
6	0.32\\
6	0.33\\
6	0.34\\
6	0.35\\
6	0.36\\
6	0.37\\
6	0.38\\
6	0.39\\
7	0.4\\
7	0.41\\
7	0.42\\
7	0.43\\
7	0.44\\
7	0.45\\
7	0.46\\
7	0.47\\
7	0.48\\
7	0.49\\
7	0.5\\
7	0.51\\
7	0.52\\
7	0.53\\
7	0.54\\
7	0.55\\
7	0.56\\
7	0.57\\
7	0.58\\
7	0.59\\
7	0.6\\
7	0.61\\
7	0.62\\
7	0.63\\
7	0.64\\
7	0.65\\
7	0.66\\
7	0.67\\
8	0.68\\
8	0.69\\
8	0.7\\
8	0.71\\
8	0.72\\
8	0.73\\
8	0.74\\
8	0.75\\
8	0.76\\
8	0.77\\
8	0.78\\
8	0.79\\
8	0.8\\
8	0.81\\
8	0.82\\
8	0.83\\
8	0.84\\
9	0.85\\
9	0.86\\
9	0.87\\
9	0.88\\
9	0.89\\
9	0.9\\
10	0.91\\
10	0.92\\
10	0.93\\
11	0.94\\
11	0.95\\
11	0.96\\
11	0.97\\
11	0.98\\
12	0.99\\
12	1\\
};
\addlegendentry{SDR-RS}

\addplot[color=green!70!black, opacity=0.8, line width=1.5pt]
table[row sep=crcr]{
22	0.01\\
22	0.02\\
30	0.03\\
32	0.04\\
35	0.05\\
36	0.06\\
36	0.07\\
39	0.08\\
39	0.09\\
40	0.1\\
44	0.11\\
44	0.12\\
45	0.13\\
45	0.14\\
46	0.15\\
46	0.16\\
47	0.17\\
49	0.18\\
49	0.19\\
50	0.2\\
51	0.21\\
52	0.22\\
52	0.23\\
54	0.24\\
55	0.25\\
55	0.26\\
56	0.27\\
56	0.28\\
57	0.29\\
57	0.3\\
58	0.31\\
58	0.32\\
58	0.33\\
60	0.34\\
64	0.35\\
64	0.36\\
65	0.37\\
66	0.38\\
66	0.39\\
66	0.4\\
68	0.41\\
69	0.42\\
69	0.43\\
71	0.44\\
71	0.45\\
71	0.46\\
72	0.47\\
75	0.48\\
75	0.49\\
76	0.5\\
76	0.51\\
77	0.52\\
78	0.53\\
78	0.54\\
78	0.55\\
79	0.56\\
80	0.57\\
82	0.58\\
83	0.59\\
83	0.6\\
84	0.61\\
84	0.62\\
86	0.63\\
86	0.64\\
88	0.65\\
88	0.66\\
90	0.67\\
92	0.68\\
95	0.69\\
95	0.7\\
95	0.71\\
95	0.72\\
98	0.73\\
100	0.74\\
100	0.75\\
103	0.76\\
104	0.77\\
107	0.78\\
109	0.79\\
112	0.8\\
112	0.81\\
114	0.82\\
117	0.83\\
119	0.84\\
123	0.85\\
129	0.86\\
132	0.87\\
140	0.88\\
142	0.89\\
148	0.9\\
151	0.91\\
157	0.92\\
159	0.93\\
166	0.94\\
167	0.95\\
170	0.96\\
202	0.97\\
204	0.98\\
245	0.99\\
320	1\\
};
\addlegendentry{SIWMMSE-RS}

\end{axis}

\end{tikzpicture}%

%% file: CDF_Runtime_M=16_Tdl=3_Pdl=20dB_alpha=0.5.tex
%
%
\definecolor{mycolor1}{rgb}{0.00000,0.44700,0.74100}%
\definecolor{mycolor2}{rgb}{0.85000,0.32500,0.09800}%
\definecolor{mycolor3}{rgb}{0.92900,0.69400,0.12500}%
\begin{tikzpicture}

\begin{axis}[%
width=3.721in,
height=3.566in,
at={(0.758in,0.481in)},
scale only axis,
xmin=0,
xmax=300,
xlabel style={font=\color{white!15!black},font=\LARGE},
xlabel={Run time [s]},
ymin=0,
ymax=1,
yticklabel style={font=\LARGE},
xticklabel style={font=\LARGE},
ylabel style={font=\color{white!15!black}, font=\huge},
ylabel={CDF},
axis background/.style={fill=white},
xtick={0,100,200,300},
ytick={0,0.5,1},
xmajorgrids,
ymajorgrids,
legend style={at={(0.48,0.5)}, anchor=south west, legend cell align=left, align=left, draw=white!15!black, font=\LARGE}
]
\addplot  [color=red, line width=2.5pt]
  table[row sep=crcr]{%
0.013402	0.01\\
0.014728	0.02\\
0.014902	0.03\\
0.015957	0.04\\
0.01781	0.05\\
0.017847	0.06\\
0.0184	0.07\\
0.019009	0.08\\
0.019127	0.09\\
0.019264	0.1\\
0.020172	0.11\\
0.020393	0.12\\
0.020816	0.13\\
0.021448	0.14\\
0.021706	0.15\\
0.022459	0.16\\
0.023092	0.17\\
0.024644	0.18\\
0.024839	0.19\\
0.025179	0.2\\
0.025804	0.21\\
0.026363	0.22\\
0.026616	0.23\\
0.027923	0.24\\
0.028401	0.25\\
0.028623	0.26\\
0.028851	0.27\\
0.02893	0.28\\
0.029343	0.29\\
0.03251	0.3\\
0.03371	0.31\\
0.034215	0.32\\
0.034442	0.33\\
0.034685	0.34\\
0.034784	0.35\\
0.034887	0.36\\
0.037097	0.37\\
0.038242	0.38\\
0.038268	0.39\\
0.038473	0.4\\
0.038536	0.41\\
0.038687	0.42\\
0.04068	0.43\\
0.040747	0.44\\
0.042132	0.45\\
0.04297	0.46\\
0.043442	0.47\\
0.044046	0.48\\
0.047984	0.49\\
0.048159	0.5\\
0.048358	0.51\\
0.049163	0.52\\
0.050793	0.53\\
0.050928	0.54\\
0.053325	0.55\\
0.055248	0.56\\
0.055572	0.57\\
0.058408	0.58\\
0.058699	0.59\\
0.061059	0.6\\
0.062303	0.61\\
0.062828	0.62\\
0.066231	0.63\\
0.066295	0.64\\
0.068157	0.65\\
0.068817	0.66\\
0.069574	0.67\\
0.071001	0.68\\
0.072336	0.69\\
0.073736	0.7\\
0.074406	0.71\\
0.076358	0.72\\
0.077781	0.73\\
0.086542	0.74\\
0.095002	0.75\\
0.098014	0.76\\
0.107969	0.77\\
0.110009	0.78\\
0.114838	0.79\\
0.132319	0.8\\
0.132371	0.81\\
0.13407	0.82\\
0.139837	0.83\\
0.14039	0.84\\
0.145623	0.85\\
0.146145	0.86\\
0.149779	0.87\\
0.154503	0.88\\
0.1761	0.89\\
0.178987	0.9\\
0.198245	0.91\\
0.200727	0.92\\
0.202083	0.93\\
0.20719	0.94\\
0.214725	0.95\\
0.245252	0.96\\
0.264912	0.97\\
0.278954	0.98\\
0.28092	0.99\\
0.48473	1\\
};
\addlegendentry{\textbf{AWAMSE-RS}}

\addplot [color=blue,opacity=0.8, line width=1.5pt]
  table[row sep=crcr]{%
55.510022	0.01\\
58.189449	0.02\\
60.190168	0.03\\
61.981339	0.04\\
62.169267	0.05\\
65.846111	0.06\\
67.41649	0.07\\
67.543444	0.08\\
68.350393	0.09\\
68.749059	0.1\\
69.43516	0.11\\
69.496496	0.12\\
69.598911	0.13\\
69.620848	0.14\\
69.647388	0.15\\
70.370328	0.16\\
70.743419	0.17\\
70.833465	0.18\\
70.918269	0.19\\
71.031499	0.2\\
71.043922	0.21\\
71.437207	0.22\\
72.227964	0.23\\
72.391218	0.24\\
72.781297	0.25\\
72.814298	0.26\\
73.044633	0.27\\
73.239262	0.28\\
73.751139	0.29\\
73.929591	0.3\\
74.029252	0.31\\
74.160886	0.32\\
74.842987	0.33\\
74.88983	0.34\\
74.980844	0.35\\
75.122207	0.36\\
75.410618	0.37\\
75.422948	0.38\\
75.5942	0.39\\
76.235414	0.4\\
76.738213	0.41\\
77.120074	0.42\\
77.23596	0.43\\
77.258175	0.44\\
77.941526	0.45\\
78.030503	0.46\\
78.420655	0.47\\
78.530533	0.48\\
79.362151	0.49\\
79.406211	0.5\\
80.175433	0.51\\
80.18963	0.52\\
81.503772	0.53\\
81.685048	0.54\\
82.695951	0.55\\
82.860274	0.56\\
83.698626	0.57\\
84.100225	0.58\\
84.716195	0.59\\
85.010065	0.6\\
85.80122	0.61\\
86.128549	0.62\\
86.205185	0.63\\
86.407984	0.64\\
87.436629	0.65\\
87.880328	0.66\\
88.207944	0.67\\
89.233585	0.68\\
90.899965	0.69\\
91.019666	0.7\\
91.023314	0.71\\
93.951852	0.72\\
93.991963	0.73\\
94.160598	0.74\\
94.624315	0.75\\
96.210292	0.76\\
96.596527	0.77\\
97.004249	0.78\\
97.636224	0.79\\
97.917313	0.8\\
99.042465	0.81\\
99.39838	0.82\\
100.238385	0.83\\
101.368825	0.84\\
101.708756	0.85\\
101.730969	0.86\\
102.749352	0.87\\
105.146259	0.88\\
110.396705	0.89\\
112.438798	0.9\\
115.267417	0.91\\
118.677602	0.92\\
119.440744	0.93\\
122.489007	0.94\\
126.052927	0.95\\
127.649597	0.96\\
131.694391	0.97\\
140.78738	0.98\\
143.757436	0.99\\
145.675324	1\\
};
\addlegendentry{SDR-RS}

\addplot [color=green!70!black, opacity=0.8, line width=1.5pt]
  table[row sep=crcr]{%
19.244481	0.01\\
19.815485	0.02\\
26.380244	0.03\\
29.830638	0.04\\
30.84204	0.05\\
31.480251	0.06\\
32.996276	0.07\\
34.199785	0.08\\
34.428261	0.09\\
35.140507	0.1\\
38.334744	0.11\\
38.500727	0.12\\
39.422019	0.13\\
40.992197	0.14\\
41.826516	0.15\\
42.329311	0.16\\
42.533674	0.17\\
43.103255	0.18\\
43.297	0.19\\
43.919151	0.2\\
45.315952	0.21\\
45.607961	0.22\\
46.943198	0.23\\
47.84759	0.24\\
47.91828	0.25\\
48.686556	0.26\\
49.02917	0.27\\
50.041136	0.28\\
50.342919	0.29\\
50.664542	0.3\\
51.15038	0.31\\
51.237421	0.32\\
51.638522	0.33\\
53.112748	0.34\\
56.24653	0.35\\
57.25721	0.36\\
57.650584	0.37\\
59.043173	0.38\\
59.294527	0.39\\
59.572326	0.4\\
60.351881	0.41\\
61.198541	0.42\\
61.563977	0.43\\
62.008347	0.44\\
63.651776	0.45\\
64.34374	0.46\\
65.25783	0.47\\
66.649159	0.48\\
67.011259	0.49\\
67.198531	0.5\\
67.9904	0.51\\
68.392643	0.52\\
68.744205	0.53\\
69.172743	0.54\\
71.080017	0.55\\
71.254838	0.56\\
71.880474	0.57\\
73.073562	0.58\\
73.147268	0.59\\
74.284517	0.6\\
75.952661	0.61\\
76.117514	0.62\\
76.623663	0.63\\
76.866604	0.64\\
77.279693	0.65\\
77.653846	0.66\\
81.840794	0.67\\
82.337426	0.68\\
82.533619	0.69\\
83.423255	0.7\\
85.026207	0.71\\
85.912495	0.72\\
88.794934	0.73\\
89.133231	0.74\\
89.484466	0.75\\
93.609672	0.76\\
94.497457	0.77\\
96.939899	0.78\\
97.545881	0.79\\
97.761053	0.8\\
100.137237	0.81\\
101.987675	0.82\\
104.4955	0.83\\
107.921801	0.84\\
109.511278	0.85\\
110.208828	0.86\\
113.031159	0.87\\
123.477679	0.88\\
124.566588	0.89\\
132.242559	0.9\\
132.937374	0.91\\
135.684812	0.92\\
142.050995	0.93\\
142.997714	0.94\\
145.554787	0.95\\
148.806926	0.96\\
177.163307	0.97\\
184.544284	0.98\\
214.121823	0.99\\
293.03181	1\\
};
\addlegendentry{SIWMMSE-RS}

\end{axis}

\end{tikzpicture}%

%% file: SR_M=16_K=5_Tdl=2_alpha=0.5_Quadriga.tex
%
%
\definecolor{mycolor1}{rgb}{0.07451,0.62353,1.00000}
\begin{tikzpicture}

\begin{axis}[%
width=4.521in,
height=3.566in,
at={(0.758in,0.481in)},
scale only axis,
xmin=0,
xmax=40,
xlabel style={font=\color{white!15!black},font=\LARGE},
xlabel={$\text{P}_{\text{dl}}\text{ [dB]}$},
ymin=0,
ymax=30,
ylabel style={font=\color{white!15!black}, font=\LARGE},
ylabel={$R_\text{sum}$ [bpcu]},
axis background/.style={fill=white},
xtick={10,20,30,40},
ytick={0,10,20,30},
xmajorgrids,
ymajorgrids,
yticklabel style={font=\Large},
xticklabel style={font=\Large},
legend style={at={(0.15,0.6)}, anchor=south west, legend cell align=left, align=left, draw=white!15!black, font=\Large}
]

\addplot [color=green!70!black, opacity=0.8, line width=1pt, mark=o, mark options={solid},mark size=4pt]
  table[row sep=crcr]{%
0	2.73381275470539\\
10	7.14894011800016\\
20	13.2812808482852\\
30	19.5812521959138\\
40	26.0849130454893\\
};
\addlegendentry{SIWMMSE-RS}

\addplot[color=blue,opacity=0.8, line width=1pt, mark=+, mark options={solid, blue}, mark size=4pt]
  table[row sep=crcr]{%
0	2.76831656608355\\
10	7.1158297102526\\
20	12.9060352556727\\
30	19.508649059555\\
40	26.0135185264882\\
};
\addlegendentry{SDR-RS}

\addplot [color=red, line width=1.5pt, mark=triangle, mark options={solid, rotate=270, red},mark size=4pt]
  table[row sep=crcr]{%
0	2.7572614867244\\
10	7.18252566482783\\
20	13.1414710253609\\
30	19.6576579282857\\
40	26.105448165639\\
};
\addlegendentry{\textbf{AWAMSE-RS}}

\addplot [color=red!40!white,line width=1.5pt, mark=triangle, mark options={solid ,rotate=270}, dashed, mark size=4pt]
  table[row sep=crcr]{%
0	2.69592795030321\\
10	6.54245331501811\\
20	7.3441636504084\\
30	7.75210754671082\\
40	8.86425979259009\\
};
\addlegendentry{AWAMSE-noRS}

\addplot [color=mycolor1,line width=1.5pt, mark=square, mark options={solid}, dotted, mark size=3]
  table[row sep=crcr]{%
0	2.00330042695999\\
10	4.08405979196328\\
20	4.99653769113137\\
30	5.14876203435989\\
40	5.16900373120282\\
};
\addlegendentry{MMSE}

\end{axis}
\end{tikzpicture}%

%% file: PowerAlloc_M=16_Tdl=2_Pdl=40dB_alpha=0.5.tex
%
%
\definecolor{mycolor1}{rgb}{0.07451,0.62353,1.00000}
\begin{tikzpicture}

\begin{axis}[%
width=3.721in,
height=3in,
at={(0.758in,0.481in)},
scale only axis,
ybar=2.5pt,
bar width=.3cm,
xmin=0.514285714285714,
xmax=6.48571428571429,
xtick={1,2,3,4,5,6},
yticklabel style={font=\LARGE},
xticklabel style={font=\LARGE},
xticklabels={Com, UE1, UE2, UE3, UE4, UE5},
ymin=0,
ymax=1,
ytick={0,.25,.5,.75,1},
yticklabels={0,,0.5,,1},
ylabel style={font=\color{white!15!black}, font=\huge},
ylabel={Power fraction},
axis background/.style={fill=white},
xmajorgrids,
ymajorgrids,
legend style={at={(0.25,0.7)}, anchor=south west, legend cell align=left, align=left, draw=white!15!black, font=\LARGE},
legend image code/.code={
        \draw [#1] (0cm,-0.2cm) rectangle (0.2cm,0.3cm); }
]

\addplot[black,fill=green!70!black, draw=black, opacity=0.5] coordinates {
(1,	0.622160950836062) (2,0.170568810327004)  (3,3.093643054671235e-49) (4,3.091582012534090e-49) (5,0.207262768014144)  (6,3.097250947255805e-49)
  };

\addplot[black,fill=blue, opacity=0.5, draw=black] coordinates {
(1,	0.390852168945329) (2,0.302296409432954) (3,5.216445619634179e-12) (4,3.816245613337704e-12)  (5,0.306851421602392) (6,1.029283630270364e-11)
  };

\addplot[black,fill=red, draw=black] coordinates {
(1,	0.557918949895753) (2,0.198400657420611) (3, 8.498305407311155e-195) (4,5.493354021722691e-230)  (5,0.243680392683637) (6,1.752243449629729e-206)
  };

\legend{SIWMMSE-RS, AO-RS, \textbf{AWAMSE-RS}}

\end{axis}

\end{tikzpicture}%

%% file: PowerAlloc_M=16_Tdl=3_Pdl=40dB_alpha=0.5.tex
%
%
\definecolor{mycolor1}{rgb}{0.07451,0.62353,1.00000}
\begin{tikzpicture}

\begin{axis}[%
width=3.721in,
height=3in,
at={(0.758in,0.481in)},
scale only axis,
ybar=2.5pt,
bar width=.3cm,
xmin=0.514285714285714,
xmax=6.48571428571429,
xtick={1,2,3,4,5,6},
yticklabel style={font=\LARGE},
xticklabel style={font=\LARGE},
xticklabels={Com, UE1, UE2, UE3, UE4, UE5},
ymin=0,
ymax=1,
ytick={0,.25,.5,.75,1},
yticklabels={0,,0.5,,1},
ylabel style={font=\color{white!15!black}, font=\huge},
ylabel={Power fraction},
axis background/.style={fill=white},
xmajorgrids,
ymajorgrids,
legend style={at={(0.25,0.7)}, anchor=south west, legend cell align=left, align=left, draw=white!15!black, font=\LARGE},
legend image code/.code={
        \draw [#1] (0cm,-0.2cm) rectangle (0.2cm,0.3cm); }
]

\addplot[black,fill=green!70!black, draw=black, opacity=0.5] coordinates {
(1,	0.574406428520847) (2,0.073692408560262)  (3,4.695904755893958e-50) (4,4.694205076971481e-50) (5,0.173528358829773)  (6,0.178372151482446)
  };

\addplot[black,fill=blue, opacity=0.5, draw=black] coordinates {
(1,	0.390058123520769) (2,0.182964002516542) (3, 8.255516380155630e-14) (4, 1.121810803908535e-13)  (5,0.199005681702731) (6,0.227972192259764)
  };

\addplot[black,fill=red, draw=black] coordinates {
(1,	0.322300990156079) (2,0.102860000444109) (3, 1.238522617154924e-60) (4, 3.124203391097174e-61)  (5,0.241438859345504) (6,0.333400150054308)
  };

\legend{SIWMMSE-RS, AO-RS, \textbf{AWAMSE-RS}}

\end{axis}

\end{tikzpicture}%

%% file: main.bbl
\begin{thebibliography}{10}
	
	\bibitem{RSMA-6G}
	O.~Dizdar, Y.~Mao, W.~Han, and B.~Clerckx.
	\newblock {Rate-Splitting Multiple Access: A New Frontier for the PHY Layer of
		6G}.
	\newblock In {\em 2020 IEEE 92nd Vehicular Technology Conference
		(VTC2020-Fall)}, pages 1--7, 2020.
	
	\bibitem{RSMA-Fundamentals}
	Y.~Mao, O.~Dizdar, B.~Clerckx, R.~Schober, P.~Popovski, and H.~V. Poor.
	\newblock {Rate-Splitting Multiple Access: Fundamentals, Survey, and Future
		Research Trends}.
	\newblock {\em IEEE Communications Surveys $\&$ Tutorials}, 24(4):2073--2126,
	2022.
	
	\bibitem{RS-PHY}
	B.~Clerckx, H.~Joudeh, C.~Hao, M.~Dai, and B.~Rassouli.
	\newblock {Rate splitting for MIMO wireless networks: a promising PHY-layer
		strategy for LTE evolution}.
	\newblock {\em IEEE Communications Magazine}, 54(5):98--105, 2016.
	
	\bibitem{RS-PHY2}
	O.~Dizdar, Y.~Mao, W.~Han, and B.~Clerckx.
	\newblock {Rate-Splitting Multiple Access: A New Frontier for the PHY Layer of
		6G}.
	\newblock In {\em 2020 IEEE 92nd Vehicular Technology Conference
		(VTC2020-Fall)}, pages 1--7, 2020.
	
	\bibitem{RS-perfectCSI1}
	A.~R. Flores, R.~C. de~Lamare, and B.~Clerckx.
	\newblock {Linear Precoding and Stream Combining for Rate Splitting in
		Multiuser MIMO Systems}.
	\newblock {\em IEEE Communications Letters}, 24(4):890--894, 2020.
	
	\bibitem{RS-perfectCSI2}
	G.~Zhou, Y.~Mao, and B.~Clerckx.
	\newblock {Rate-Splitting Multiple Access for Multi-Antenna Downlink
		Communication Systems: Spectral and Energy Efficiency Tradeoff}.
	\newblock {\em IEEE Transactions on Wireless Communications}, 21(7):4816--4828,
	2022.
	
	\bibitem{RS-imperfectCSI1}
	J.~An, O.~Dizdar, B.~Clerckx, and W.~Shin.
	\newblock {Rate-Splitting Multiple Access for Multi-Antenna Broadcast Channel
		with Imperfect CSIT and CSIR}.
	\newblock In {\em IEEE Annu. Symp. Pers. Indoor Mobile Radio Commun. (PIMRC)},
	pages 1--7, 2020.
	
	\bibitem{RS-imperfectCSI2}
	M.~Dai, B.~Clerckx, D.~Gesbert, and G.~Caire.
	\newblock {A Rate Splitting Strategy for Massive MIMO With Imperfect CSIT}.
	\newblock {\em IEEE Transactions on Wireless Communications}, 15(7):4611--4624,
	2016.
	
	\bibitem{ICC24}
	D.~Ben~Amor, M.~Joham, and W.~Utschick.
	\newblock {Highly Accelerated Weighted MMSE Algorithms for Designing Precoders
		in FDD Systems with Incomplete CSI}.
	\newblock 2023.
	\newblock \url{https://arxiv.org/abs/2312.01888}.
	
	\bibitem{Clerckx_RS3}
	H.~Joudeh and B.~Clerckx.
	\newblock {Robust Transmission in Downlink Multiuser {MISO} Systems: A
		Rate-Splitting Approach}.
	\newblock {\em IEEE Transactions on Signal Processing}, 64(23):6227--6242,
	2016.
	
	\bibitem{Hasssibi_Hochwald}
	B.~Hassibi and B.M. Hochwald.
	\newblock {How Much Training is Needed in Multiple-Antenna Wireless Links?}
	\newblock {\em {IEEE Transactions on Information Theory}}, 49(4):951--963,
	2003.
	
	\bibitem{SpecMang_Luo}
	Z.~Luo and S.~Zhang.
	\newblock {Dynamic Spectrum Management: Complexity and Duality}.
	\newblock {\em IEEE Journal of Selected Topics in Signal Processing},
	2(1):57--73, 2008.
	
	\bibitem{WienerFilter}
	M.~Joham, K.~Kusume, M.H. Gzara, W.~Utschick, and J.A. Nossek.
	\newblock {Transmit Wiener filter for the downlink of TDDDS-CDMA systems}.
	\newblock In {\em IEEE Seventh International Symposium on Spread Spectrum
		Techniques and Applications,}, volume~1, pages 9--13 vol.1, 2002.
	
	\bibitem{RethinkWMMSE}
	X.~Zhao, S.~Lu, Q.~Shi, and Z.~Luo.
	\newblock {Rethinking WMMSE: Can Its Complexity Scale Linearly With the Number
		of BS Antennas?}
	\newblock {\em IEEE Transactions on Signal Processing}, 71:433--446, 2023.
	
	\bibitem{GMM2}
	M.~Koller, B.~Fesl, N.~Turan, and W.~Utschick.
	\newblock {An Asymptotically MSE-Optimal Estimator Based on Gaussian Mixture
		Models}.
	\newblock {\em IEEE Transactions on Signal Processing}, 70:4109--4123, 2022.
	
	\bibitem{ZF_LS}
	D.~Ben~Amor, M.~Joham, and W.~Utschick.
	\newblock {Asymptotic Behavior of Zero-Forcing Precoding based on Imperfect
		Channel Knowledge for Massive MISO FDD Systems}.
	\newblock In {\em ICC 2023 - IEEE International Conference on Communications},
	pages 1500--1505, 2023.
	
	\bibitem{cvx}
	M.~Grant and S.~Boyd.
	\newblock {{CVX}: Matlab Software for Disciplined Convex Programming, version
		2.1}.
	\newblock \url{http://cvxr.com/cvx}, March 2014.
	
	\bibitem{MMSEprecoder}
	E.~Björnson, J.~Hoydis, and L.~Sanguinetti.
	\newblock {Massive MIMO Has Unlimited Capacity}.
	\newblock {\em IEEE Transactions on Wireless Communications}, 17(1):574--590,
	2018.
	
\end{thebibliography}
